\documentclass[aps,preprint,floats,epsf,epsfig,nofootinbib,letter]{revtex4}
\usepackage{epsfig,graphicx,color,appendix}

\begin{document}
\begin{flushright}
KIAS-P12044
\end{flushright}

\def\CP{{\it CP}~}
\def\cp{{\it CP}}

\title{\mbox{}\\[11pt]
 Recent Neutrino Data and a Realistic Tribimaximal-like Neutrino Mixing Matrix}

\author{Y. H. Ahn$^1$\footnote{Email: yhahn@kias.re.kr},
Hai-Yang Cheng$^2$\footnote{Email: phcheng@phys.sinica.edu.tw},
and Sechul Oh$^3$\footnote{Email: scohph@yonsei.ac.kr}}

\affiliation{
$^1$School of Physics, KIAS, Seoul 130-722, Korea  \\
$^2$Institute of Physics, Academia Sinica, Taipei 115, Taiwan \\
$^3$University College, Yonsei University, Incheon 406-840, Korea
}

\date{\today}

\begin{abstract}
In light of the recent neutrino experimental results from Daya Bay and RENO Collaborations, we construct a realistic tribimaximal-like Pontecorvo-Maki-Nakagawa-Sakata (PMNS) leptonic mixing matrix.
Motivated by the Qin-Ma (QM) parametrization for the quark mixing matrix in which the {\it CP}-odd phase is approximately maximal,
we propose a simple {\it ansatz} for the charged lepton mixing matrix, namely, it has the QM-like parametrization, and assume the tribimaximal mixing (TBM) pattern for the neutrino mixing matrix.
The deviation of the leptonic mixing matrix from the TBM one is then systematically studied. While the deviation of the solar and atmospheric neutrino mixing angles from the corresponding TBM values, i.e. $\sin^2\theta_{12}=1/3$ and $\sin^2\theta_{23}=1/2$, is fairly small,
we find a nonvanishing reactor mixing angle given by $\sin\theta_{13}\approx \lambda/\sqrt{2}$ ($\lambda\approx 0.22$ being the Cabibbo angle). Specifically, we obtain $\theta_{13}\simeq 9.2^{\circ}$ and
$\delta_{CP} \simeq \delta_{\rm QM} \simeq {\cal O}( 90^{\circ})$.
Furthermore, we show that the leptonic \CP violation characterized by the Jarlskog
invariant is $|J^{\ell}_{CP}|\simeq \lambda/6$, which could be tested in the future experiments
such as the upcoming long baseline neutrino oscillation ones.

\end{abstract}

\maketitle %
\section{Introduction}
Recent analyses of the neutrino oscillation data \cite{nudata,Fogli:2012ua} indicate that the tribimaximal mixing (TBM)
pattern for three flavors of neutrinos \cite{tribi} can be regarded as the zeroth order leptonic mixing matrix
 \begin{eqnarray}
   U_{\rm TB}={\left(\begin{array}{ccc}
   \frac{2}{\sqrt{6}} & \frac{1}{\sqrt{3}} & 0 \\
   -\frac{1}{\sqrt{6}} & \frac{1}{\sqrt{3}} & -\frac{1}{\sqrt{2}}  \\
   -\frac{1}{\sqrt{6}} & \frac{1}{\sqrt{3}} & \frac{1}{\sqrt{2}}
   \end{array}\right)}P_{\nu} \ ,
   \label{TBM}
 \end{eqnarray}
where $P_{\nu}={\rm Diag}(e^{i\delta_{1}},e^{i\delta_{2}},1)$ is a diagonal matrix of phases for the Majorana
neutrino. However, properties related to the leptonic \CP violation remain completely unknown yet. The large values of
the solar and atmospheric mixing angles, which may be suggestive of a new flavor symmetry in the lepton sector, are entriely different from the
quark mixing ones. The structure of both charged lepton and neutrino mass matrices could be dedicated by
a flavor symmetry, for example, the $A_4$ discrete symmetry, which will tell us something about the charged fermion and neutrino mixings.  If there exists such a flavor symmetry in nature, the TBM pattern for the neutrino mixing matrix may come out in a natural way.
It is well known that there are no sizable effects on the observables from the renormalization group running for the hierarchical mass spectrum in the standard model~\cite{RG}.~\footnote{This may not be true beyond the standard model. For example, for quasi-degenerate neutrinos and large $\tan\beta$ in the minimal supersymmetric model, all three mixing angles may change significantly~\cite{Dighe:2006sr}.}
Hence, corrections to the tribimaximal  neutrino mixing from renormalization
group effects running down from the seesaw scale are negligible in the standard model.

The so-called PMNS (Pontecorvo-Maki-Nakagawa-Sakata) leptonic mixing matrix depends generally on the charged lepton sector whose diagonalization leads to a charged lepton mixing matrix $V^{\ell}_{L}$ which should be combined with the neutrino mixing matrix $U_\nu$; that is,
\begin{eqnarray}
  U_{\rm PMNS}= V^{\ell\dag}_{L}U_\nu \ .
 \label{mixing relation}
\end{eqnarray}
In the charged fermion (quarks and charged leptons) sector, there is a qualitative feature which distinguishes the neutrino sector from the charged fermion one. The mass spectrum of the charged leptons exhibits a similar hierarchical pattern as that of the down-type quarks, unlike that of the up-type quarks which shows a much stronger hierarchical pattern. For example, in terms of the Cabbibo angle $\lambda \equiv \sin\theta_{\rm C} \approx |V_{us}|$, the
fermion masses scale as ~$(m_{e},m_{\mu}) \approx (\lambda^{5},\lambda^{2})~ m_{\tau}$,~$(m_{d},m_{s}) \approx (\lambda^{4},\lambda^{2})~ m_{b}$~ and
~$(m_{u},m_{c}) \approx (\lambda^{8},\lambda^{4})~ m_{t}$.  This  may lead to two implications: (i) the Cabibbo-Kobayashi-Maskawa (CKM) matrix \cite{CKM} is mainly governed by the down-type quark mixing matrix, and (ii) the charged lepton mixing matrix is similar to that of the down-type quark one. Therefore, we shall assume that (i) $V_{\rm CKM}=V^{d\dag}_{L}$ and $V^{u}_{L}=\mathbf{1}$, where $V^{d}_{L}~(V^{u}_{L})$ is associated with the diagonalization of the down-type (up-type) quark mass matrix and $\mathbf{1}$ is a $3\times3$ unit matrix, and (ii) the charged lepton mixing matrix $V^{\ell \dag}_{L}$ has the same structure as the CKM matrix, $V^{\ell \dag}_{L}=V_{\rm CKM}$.

Very recently, a non-vanishing mixing angle $\theta_{13}$ has been reported firstly from Daya Bay and RENO Collaborations~\cite{An:2012eh,Ahn:2012nd} with the results given by
 \begin{eqnarray}
  \sin^2 2\theta_{13}=0.092\pm 0.016(stat)\pm 0.005(syst)
 \end{eqnarray}
 and
 \begin{eqnarray}
  \sin^2 2\theta_{13}=0.113\pm 0.013(stat)\pm 0.019(syst)~,
 \end{eqnarray}
respectively. These results are in good agreement with the previous data from the T2K, MINOS and Double Chooz Collaborations~\cite{Data}. The experimental results of the non-zero $\theta_{13}$ indicate that the TBM pattern for the neutrino mixing should be modified. Moreover, properties related to the leptonic CP violation remain completely unknown yet.

In this work, we shall assume a neutrino mixing matrix in the TBM form,
\begin{eqnarray}  \label{Urelation}
  U_{\nu}= U_{\rm TB} \ .
\end{eqnarray}
We will neglect possible corrections to $U_{\rm TB}$ from higher dimensional operators and from renormalization group effects.
Then we make a simple {\it ansatz} on the charged lepton mixing matrix $V_L^\ell$, namely, we assume that $V_L^\ell$ has the same structure as the Qin-Ma (QM) parametrization of the quark mixing matrix, which is a Wolfenstein-like parametrization and can be expanded in terms of the small parameter $\lambda$\,. Unlike the original Wolfenstein parametrization, the QM one has the advantage that its {\it CP}-odd phase $\delta$ is manifested in the parametrization and approximately maximal, i.e. $\delta\sim 90^\circ$. As we shall see blow, this is crucial for a viable neutrino phenomenology. It turns out that the PMNS leptonic mixing matrix is identical to the TBM matrix plus some small corrections arising from the charged mixing matrix $V_L^\ell$  expanded in terms of the small parameter $\lambda$.~\footnote{There are several papers~\cite{BM} implemented in $U_{\rm PMNS}= U_{\rm BM}+{\rm corrections~in~powers~of}~\lambda$, where the bimaximal matrix $U_{\rm BM}$ leads to $\theta^{\nu}_{23}=\pi/4,~\theta^{\nu}_{12}=\pi/4$ and $\theta^{\nu}_{13}=0$.} Schematically,
\begin{eqnarray}
  U_{\rm PMNS}= U_{\rm TB}+~{\rm corrections~in~powers~of}~\lambda \ .
\end{eqnarray}
Consequently, not only
the solar and atmospheric mixing angles given by the TBM remain valid but also the reactor mixing angle and the Dirac phase can be deduced from the above consideration.

The paper is organized as follows. In Sec. II we discuss the parameterizations of quark and lepton mixing matrices and pick up the one suitable for our purpose in this work. After making an ansatz on the charged lepton mixing matrix we study the low-energy neutrino phenomenology and emphasize the new results on the reactor neutrino mixing angle and the {\it CP}-odd phase in Sec. III. Our conclusions are summarized in Sec. IV.

\section{Lepton and quark mixing}

In the weak eigenstate basis, the Lagrangian relevant to the lepton sector reads
 \begin{eqnarray}
 -{\cal L} &=& \frac{1}{2}\overline{\nu_{L}}m_{\nu}(\nu_{L})^c+\overline{\ell_{L}}m_{\ell}\ell_{R}
 +\frac{g}{\sqrt{2}}W^{-}_{\mu}\overline{\ell_{L}}\gamma^{\mu}\nu_{L} + {\rm H.c.} ~
 \label{lagrangianA}
 \end{eqnarray}
When diagonalizing the neutrino and charged lepton mass matrices,
 $U^{\dag}_{\rm TB}m_{\nu}U^{\ast}_{\rm TB}={\rm Diag}(m_{1},m_{2},m_{3})$ and $V^{\ell\dag}_{L}m_{\ell}V^{\ell}_{R} = {\rm Diag}(m_{e},m_{\mu},m_{\tau})$,
we can rotate the fermion fields from the weak eigenstates to the mass eigenstates, $\nu_{L}\rightarrow U^{\dag}_\nu\nu_{L},~\ell_{L}\rightarrow V^{\ell ~\dag}_{L}\ell_{L},~\ell_{R}\rightarrow V^{\dag}_{R}\ell_{R}$.
Then we obtain the leptonic $3\times3$ unitary mixing matrix
  $U_{\rm PMNS}= V^{\ell\dag}_{L}U_\nu$ as given in Eq. (\ref{mixing relation}) from the charged current term in Eq.~(\ref{lagrangianA}).
In the standard parametrization of the leptonic mixing matrix  $U_{\rm PMNS}$, it is expressed in terms of three mixing angles and three \cp-odd phases (one $\delta'$ for the Dirac neutrino and two for the Majorana neutrino) \footnote{
For definiteness, we shall use the Jarlskog rephasing invariant as shown in Eq.~(\ref{Jcp2}) to define the Dirac $CP$-violating phase $\delta_{CP}$. The Dirac phase defined in this manner is independent of a particular parametrization of the PMNS matrix.
In general, $\delta'$ may not be equal to $\delta_{CP}$. It shall be shown that $\delta_{CP}$ equals to the phase $\delta$ defined in Eq.~(\ref{Vl}), up to order $\lambda^3$.} \cite{PDG}
 \begin{eqnarray}
  U_{\rm PMNS}={\left(\begin{array}{ccc}
   c_{13}c_{12} & c_{13}s_{12} & s_{13}e^{-i\delta'} \\
   -c_{23}s_{12}-s_{23}c_{12}s_{13}e^{i\delta'} & c_{23}c_{12}-s_{23}s_{12}s_{13}e^{i\delta'} & s_{23}c_{13}  \\
   s_{23}s_{12}-c_{23}c_{12}s_{13}e^{i\delta'} & -s_{23}c_{12}-c_{23}s_{12}s_{13}e^{i\delta'} & c_{23}c_{13}
   \end{array}\right)}P_{\nu}~,
 \label{PMNS}
 \end{eqnarray}
where $s_{ij}\equiv \sin\theta_{ij}$ and $c_{ij}\equiv \cos\theta_{ij}$.
The current best-fit values of $\theta_{12}$, $\theta_{23}$ and $\theta_{13}$ at 1$\sigma$ (3$\sigma$) level obtained from the global analysis by Fogli {\it et al.}~\cite{Fogli:2012ua} are given by
 \begin{eqnarray}
  \theta_{12}=33.6^{\circ+1.1^{\circ}~(+3.2^{\circ})}_{~-1.0^{\circ}~(-3.1^{\circ})}~,
  ~\theta_{23}=\left\{
  \begin{array}{ll}
  38.4^{\circ+1.4^{\circ}~(+14.4^{\circ})}_{~-1.2^{\circ}~(~-3.3^{\circ})} & \hbox{NO} \\
  38.8^{\circ+2.3^{\circ}~(+15.8^{\circ})}_{~-1.3^{\circ}~(~-3.4^{\circ})} & \hbox{IO}
  \end{array}
  \right.
  ,
  ~\theta_{13}=\left\{
  \begin{array}{ll}
  8.9^{\circ+0.5^{\circ}~(+1.3^{\circ})}_{~-0.5^{\circ}~(-1.5^{\circ})} & \hbox{NO} \\
  9.0^{\circ+0.4^{\circ}~(+1.2^{\circ})}_{~-0.5^{\circ}~(-1.5^{\circ})} & \hbox{IO}
  \end{array}
  \right.,
 \label{exp}
 \end{eqnarray}
where NO and IO stand for normal mass ordering and inverted one, respectively.
The analysis by Fogli {\it et al.} includes the updated data released at the {\it Neutino~2012} Conference~\cite{Kyoto}.
However, we know nothing at all about all three $CP$-violating phases $\delta',~\delta_{1}$ and $\delta_{2}$.

In analogy to the PMNS matrix, the CKM quark mixing matrix is a product of two unitary matrices, $V_{\rm CKM}=V^{d\dag}_{L}V^{u}_{L}$, and can be expressed in terms of four independent parameters (three mixing angles and one phase). Their current best-fit values in the $1\sigma$ range read~\cite{CKMfitter}
\begin{eqnarray} \label{eq:CKqmix}
  \theta^{q}_{12} = ( 13.03 \pm 0.05 )^{\circ},
  ~~\theta^{q}_{23} = ( 2.37^{+0.03}_{-0.07} )^{\circ}, ~
  ~~\theta^{q}_{13} = ( 0.20^{+0.01}_{-0.01} )^{\circ},
  ~~\phi = ( 67.19^{+2.40}_{-1.76} )^{\circ} ~.
 \end{eqnarray}
A well-known simple parametrization of the CKM matrix introduced
by Wolfenstein \cite{Wolfenstein:1983yz} is
\begin{eqnarray} \label{Wolf}
V_{{\rm W}}=\left( \matrix{ 1-\lambda^2/2 & \lambda &
A\lambda^3(\rho-i\eta)   \cr   -\lambda & 1-\lambda^2/2   & A\lambda^2   \cr
A\lambda^3(1-\rho-i\eta) & -A\lambda^2   & 1  \cr}\right)+{\cal O}(\lambda^4).
\end{eqnarray}
Hence, the CKM matrix is a unit matrix plus corrections expanded in powers of $\lambda$.

Recently, Qin and Ma (QM)~\cite{Qin:2010hn} have advocated a new Wolfenstein-like parametrization of the quark mixing matrix
 \begin{eqnarray}
  V_{\rm QM}={\left(\begin{array}{ccc}
   1-\lambda^2/2 & \lambda & h\lambda^3 e^{-i\delta} \\
   -\lambda & 1-\lambda^2/2 & (f+h e^{-i\delta})\lambda^2 \\
   f\lambda^3 & -(f+h e^{i\delta})\lambda^2 & 1 \\
   \end{array}\right)}+{\cal O}(\lambda^{4})~,
 \label{QM}
 \end{eqnarray}
based on the triminimal expansion of the CKM matrix.~\footnote{The phrase ``triminimal" was first used in \cite{Pakvasa} to describe the deviation from the tribimaximal pattern in the lepton mixing.}
The parameters $A$, $\rho$ and $\eta$ in the Wolfenstein parametrization~\cite{Wolfenstein:1983yz} are replaced by $f$, $h$ and $\delta$ in the Qin-Ma one.  From the global fits to the quark mixing matrix given by~\cite{CKMfitter}, we obtain
 \begin{eqnarray}
 f=0.749^{+0.034}_{-0.037}\,,\quad h=0.309^{+0.017}_{-0.012}\,,\quad \lambda=0.22545\pm 0.00065\,,
 \quad \delta=(89.6^{+2.94}_{-0.86})^\circ\,.
  \label{eq:QMfh}
 \end{eqnarray}
Therefore, the \cp-odd phase is approximately maximal in the sense that $\sin\delta\approx 1$. Because of the freedom of the phase redefinition for the quark fields, we have shown in \cite{Ahn:2011it} that the QM parametrization is indeed equivalent to the Wolfenstein one~\footnote{Relations between the Wolfenstein parameters $(A,~\rho,~\eta)$ and the QM parameters $(f,~h,~\delta)$ are shown in \cite{Ahn:2011it}.}
and pointed out that
\begin{eqnarray}
\delta= \gamma +\beta = \pi -\alpha \ ,
\end{eqnarray}
where the three angles $\alpha$, $\beta$ and $\gamma$ of the unitarity triangle are defined by
\begin{eqnarray}
\alpha\equiv{\rm arg}\left(-{V_{td}V^*_{tb}\over V_{ud}V_{ub}^*}\right),~~~
\beta\equiv{\rm arg}\left(-{V_{cd}V^*_{cb}\over V_{td}V_{tb}^*}\right),~~~
\gamma\equiv{\rm arg}\left(-{V_{ud}V^*_{ub}\over V_{cd}V_{cb}^*}\right),
\end{eqnarray}
and they satisfy the relation $\alpha+\beta+\gamma=\pi$. Since $\alpha=(91.0\pm3.9)^\circ$, $\beta=(21.76^{+0.92}_{-0.82})^\circ$ and $\gamma=(67.2\pm3.0)^\circ$ inferred from the current data
\cite{CKMfitter},
the phase $\delta$ in the QM parametrization is thus very close to $90^\circ$.

The rephasing invariant Jarlskog parameter $J^{q}_{CP}$ in the quark sector is given by
 \begin{eqnarray}
  J^{q}_{CP}={\rm Im}[V_{ud}V_{tb}V^{\ast}_{ub}V^{\ast}_{td}]=hf\lambda^{6}(1-\lambda^{2}/2)\sin\delta~,
 \label{Jcp1}
 \end{eqnarray}
implying that the phase $\delta$ in Eq.~(\ref{QM}) is equal to the rephasing invariant $CP$-violating phase.
Numerically, it reads $J^{q}_{CP}\simeq0.2\,\lambda^{6}$ using Eq.~(\ref{eq:QMfh}). For our later purpose, we shall consider a particular QM parametrization obtained by rephasing $u$ and $d$ quark fields: $u \to u \,e^{i\delta}$ and $d \to d \,e^{i\delta}$
 \begin{eqnarray}
  V_{\rm QM}={\left(\begin{array}{ccc}
   1-\lambda^2/2 & \lambda e^{i\delta} & h\lambda^3 \\
   -\lambda e^{-i\delta} & 1-\lambda^2/2 & (f+h e^{-i\delta})\lambda^2 \\
   f\lambda^3 e^{-i\delta} & -(f+h e^{i\delta})\lambda^2 & 1 \\
   \end{array}\right)}+{\cal O}(\lambda^{4})~.
 \label{QM2}
 \end{eqnarray}
As we will show in the next section, it will have
very interesting implications to the lepton sector.

\section{Low energy neutrino phenomenology}
Let us now discuss the low energy neutrino phenomenology with an {\it ansatz} that the  charged lepton mixing matrix  $V^{\ell \dag}_{L}$ has the similar expression to the QM parametrization given by Eq.~(\ref{QM2}):
 \begin{eqnarray}
  V^{\ell \dag}_{L}={\left(\begin{array}{ccc}
   1-\lambda^2/2 & \lambda e^{i\delta} & h\lambda^3 \\
   -\lambda e^{-i\delta} & 1-\lambda^2/2 & (f+h e^{-i\delta})\lambda^2 \\
   f\lambda^3 e^{-i\delta} & -(f+h e^{i\delta})\lambda^2 & 1 \\
   \end{array}\right)}+{\cal O}(\lambda^{4})~,
 \label{Vl}
 \end{eqnarray}
where the parameters $\lambda,~f,~h$ and $\delta$ in the lepton sector are {\it a priori} not necessarily the same as those in the quark sector. Nevertheless, we shall assume that $\lambda$ is a small parameter and that $\delta$ is of order $90^\circ$. As we will see below, this matrix accounts for the small deviation of the PMNS matrix from the TBM pattern.

We have emphasized in \cite{Ahn:2011yj} that the phases of the off-diagonal matrix elements of $V^{\ell}_{L}$ play a key role for a viable neutrino phenomenology. Especially, we have found that the solar mixing angle $\theta_{12}$ depends strongly on the phase of the element $(V_L^\ell)_{12}$. This is the reason why we choose the particular form of  Eq.~(\ref{Vl}).
In the quark sector, there exist infinitely many  possibilities of rephasing the quark fields in the CKM matrix and physics should be independent of the phase redefinition. The reader may wonder why we do not identify $V^{\ell}_{L}$ first with the original QM parametrization in Eq.~(\ref{QM}) and then make  phase redefinition of lepton fields to get \cp-odd phases in the off-diagonal elements. The point is that the arbitrary phase matrix of the neutrino fields does not commute with the TBM matrix $U_{\rm TB}$. As a result, the charged lepton mixing matrix in Eq.~(\ref{mixing relation}) cannot be arbitrarily rephased from the neutrino fields. Therefore, in the lepton sector, this particular form of Eq.~(\ref{Vl}) for the parametrization of $V^{\ell}_{L}$ obtained by rephasing the $u$ and
$d$ quark fields in Eq. (\ref{QM}) with a physical phase $\delta$ is the only way for $V_L^{\ell}$ consistent with the
current experimental data, especially for $\sin\theta_{12}$ (see Eq. ~(\ref{Sol}) below).

With the help of Eqs.~(\ref{mixing relation}), (\ref{Urelation}) and (\ref{Vl}), the leptonic mixing matrix corrected by the contributions from $V^{\ell}_{L}$ can be written, up to order of $\lambda^{3}$, as
 \begin{eqnarray}
  U_{\rm PMNS}&=&U_{\rm TB}+{\left(\begin{array}{ccc}
   -\frac{\lambda e^{i\delta}}{\sqrt{6}}-\frac{\lambda^{2}(1+h\lambda)}{\sqrt{6}} & \frac{\lambda e^{i\delta}}{\sqrt{3}}-\frac{\lambda^{2}(1-2h\lambda)}{2\sqrt{3}} & \frac{\lambda(h\lambda^{2}-e^{i\delta})}{\sqrt{2}} \\
   -\lambda\sqrt{\frac{2}{3}} e^{-i\delta}-\frac{\lambda^2(1-2f-2he^{-i\delta})}{2\sqrt{6}} & -\frac{\lambda e^{-i\delta}}{\sqrt{3}}-\frac{\lambda^{2}(1-2f-2h\lambda e^{-i\delta})}{2\sqrt{3}} & \frac{\lambda^2(1+2f+2h e^{-i\delta})}{2\sqrt{2}} \\
   \frac{\lambda^2(f+h e^{i\delta}+2f\lambda e^{-i\delta})}{\sqrt{6}}  & -\frac{\lambda^2(f+h e^{i\delta}-f\lambda e^{-i\delta})}{\sqrt{3}} & \frac{\lambda^2(f+h e^{i\delta})}{\sqrt{2}} \\
   \end{array}\right)}P_{\nu} \nonumber\\
   &+&{\cal O}(\lambda^4) \ .
 \label{PMNS1}
 \end{eqnarray}
By rephasing the lepton and neutrino fields $e \to e \,e^{i\alpha_{1}}$,
$\mu \to \mu \,e^{i\beta_{1}}$, $\tau \to \tau \,e^{i\beta_{2}}$ and
$\nu_{2} \to \nu_{2} \,e^{i(\alpha_{1}-\alpha_{2})}$, the PMNS matrix is recast to
 \begin{eqnarray}
  U_{\rm PMNS}=
 {\left(\begin{array}{ccc}
 |U_{e1}| & |U_{e2}| & |U_{e3}|e^{-i(\alpha_{1}-\alpha_{3})} \\
 U_{\mu1}e^{-i\beta_{1}} & U_{\mu2}e^{i(\alpha_{1}-\alpha_{2}-\beta_{1})} &  |U_{\mu3}| \\
 U_{\tau1}e^{-i\beta_{2}} & U_{\tau2}e^{i(\alpha_{1}-\alpha_{2}-\beta_{2})} & |U_{\tau3}|
 \end{array}\right) P'_{\nu}}~,
 \label{PMNS2}
 \end{eqnarray}
where $U_{\alpha j}$ is an element of the PMNS matrix  with $\alpha=e,\mu,\tau$ corresponding to the lepton flavors and $j=1,2,3$ to the light neutrino mass eigenstates.
In Eq.~(\ref{PMNS2}) the phases defined as $\alpha_{1} = \arg(U_{e1})$, $\alpha_{2} = \arg(U_{e2})$, $\alpha_{3} = \arg(U_{e3})$, $\beta_{1} = \arg(U_{\mu3})$ and $\beta_{2} = \arg(U_{\tau3})$ have the expressions
 \begin{eqnarray}
  \alpha_{1}&=&\tan^{-1}\left(\frac{-\lambda\sin\delta}{2-\lambda
  \cos\delta-\lambda^{2}-h\lambda^{3}}\right)~,\qquad\beta_{1}
  =\tan^{-1}\left(\frac{h\lambda^{2}\sin\delta}{1-\lambda^{2}(f+\frac{1}{2}+\cos\delta)}\right) \ , \nonumber\\ \alpha_{2}&=&\tan^{-1}\left(\frac{\lambda\sin\delta}{1+\lambda\cos\delta
  -\frac{\lambda^{2}}{2}+h\lambda^{3}}\right)~,\qquad\beta_{2}=\tan^{-1}
  \left(\frac{h\lambda^{2}\sin\delta}{1+\lambda^{2}(f+h\cos\delta)}\right) \ , \nonumber\\
  \alpha_{3}&=&\tan^{-1}\left(\frac{-\sin\delta}{h\lambda^{2}-\cos\delta}\right)~,\qquad\qquad \qquad~ P'_{\nu}={\rm Diag}(e^{i\delta_{1}},e^{i(\delta_{2}+\alpha_{1}-\alpha_{2})},1)~.
  \label{mixing elements2}
 \end{eqnarray}
Up to the order of $\lambda^{3}$,  the elements of $U_{\rm PMNS}$ are found to be
 \begin{eqnarray}
  |U_{e1}|&=& \sqrt{\frac{2}{3}}\left(1-\frac{\lambda\cos\delta}{2}-\frac{\lambda^{2}
  (3+\cos^{2}\delta)}{8}+\frac{\lambda^{3}}{16}(\cos\delta-8h-\cos^{3}\delta)\right)~,\nonumber\\
  |U_{e2}|&=& \frac{1}{\sqrt{3}}\left(1+\lambda\cos\delta-\frac{\lambda^{2}}{2}\cos^{2}
  \delta+\frac{\lambda^{3}}{2}(2h-\cos\delta+\cos^{3}\delta)\right)~,\nonumber\\
  |U_{e3}|&=& \frac{\lambda}{\sqrt{2}}(1-h\lambda^{2}\cos\delta)~,\nonumber\\
  U_{\mu1}&=& \frac{-1}{\sqrt{6}}\left(1+2\lambda e^{-i\delta}-\frac{\lambda^{2}}{2}(1+2f+2he^{-i\delta})\right)~,\nonumber\\
  U_{\mu2}&=& \frac{1}{\sqrt{3}}\left(1-\lambda e^{-i\delta}-\frac{\lambda^{2}}{2}(1-2f-2he^{-i\delta})\right)~,\nonumber\\
  |U_{\mu3}| &=& \frac{1}{\sqrt{2}}\left(1-\frac{\lambda^{2}}{2}(1+2f+2h\cos\delta)\right)~,\quad
  U_{\tau1}= \frac{-1}{\sqrt{6}}\left(1-\lambda^{2}(f+he^{i\delta})-2f\lambda^{3}e^{-i\delta}\right)~,\nonumber\\
  U_{\tau2}&=& \frac{1}{\sqrt{3}}\left(1-\lambda^{2}(f+he^{i\delta})+f\lambda^{3}e^{-i\delta}\right)~,~
  |U_{\tau3}|= \frac{1}{\sqrt{2}}\left(1+\lambda^{2}(f+h\cos\delta)\right)~.
  \label{mixing elements1}
 \end{eqnarray}
From Eq.~(\ref{PMNS2}), the neutrino mixing parameters can be displayed as
  \begin{eqnarray}
  \sin^{2}\theta_{12}&=&\frac{|U_{e2}|^{2}}{1-|U_{e3}|^{2}}~,\qquad\qquad\quad
   \sin^{2}\theta_{23}=\frac{|U_{\mu3}|^{2}}{1-|U_{e3}|^{2}}~,\nonumber\\
  \sin\theta_{13}&=&|U_{e3}|~.
 \label{mixing1}
 \end{eqnarray}

From Eq. (\ref{mixing elements1}), the solar neutrino mixing angle $\theta_{12}$ can be approximated, up to order $\lambda^3$, as
 \begin{eqnarray}
  \sin^{2}\theta_{12}\simeq\frac{1}{3}\left(1+2\lambda\cos\delta+\frac{\lambda^2}{2}+2h\lambda^3\right)~,
 \label{Sol}
 \end{eqnarray}
which indicates, interestingly enough, a tiny deviation from $\sin^{2}\theta_{12}=1/3$ when $\cos\delta$ approaches to zero. Since it is the first column of $V_L^\ell$ that makes the major contribution to $\sin^{2}\theta_{12}$, this explains why we need a phase of order $90^\circ$ for the element  $(V_L^\ell)_{21}$.~\footnote{In \cite{Ahn:2011yj} we have considered three different scenarios for the matrix $V_L^\ell$. We obtained the constraint $0.17\leq \cos\delta \leq 0.64$ in two of the scenarios in order to satisfy the quark-lepton complementarity (QLC) relations $\theta_{12}+\theta^q_{12}=\pi/4$ and $\theta_{23}+\theta^q_{23}=\pi/4$. In this work, we will not impose these QLC relations from the outset.
}
Likewise, the atmospheric neutrino mixing angle $\theta_{23}$ comes out as
 \begin{eqnarray}
   \sin^{2}\theta_{23}\simeq\frac{1}{2}\left(1-\frac{\lambda^{2}}{2}(4f+4h\cos\delta+1)\right)~,
 \label{Atm}
 \end{eqnarray}
which shows a very small deviation from the TBM angle $\sin^2\theta_{23}= { 1/2}$. The reactor mixing angle $\theta_{13}$ can be obtained by
 \begin{eqnarray} \label{Diracphase}
  \sin\theta_{13}&=&\frac{\lambda}{\sqrt{2}}(1-h\lambda^{2}\cos\delta)~.
 \label{Reactor}
 \end{eqnarray}
Thus, we have a nonvanishing large $\theta_{13}$.

Leptonic \CP violation at low energies could be detected through neutrino oscillations that are sensitive to the Dirac \cp-phase, but insensitive to the Majorana phases. The Jarlskog invariant for the lepton sector has the expression
 \begin{eqnarray}
  J^{\ell}_{CP}\equiv{\rm Im}[U_{e1}U_{\mu2}U^{\ast}_{e2}U^{\ast}_{\mu1}]=-\frac{\lambda\sin\delta}{6}
  \left(1-\frac{\lambda^{2}}{2}\right)+{\cal O}(\lambda^4)~.
 \label{Jcp2}
 \end{eqnarray}
This shows that up to order $\lambda^3$, the rephasing invariant Dirac $CP$-violating phase $\delta_{CP}$ equals to the phase $\delta$ introduced in Eq.~(\ref{Vl}): {\it i.e.,} $\delta_{CP} \simeq \delta$.
Also, the above relation is approximated as $J^{\ell}_{CP}\simeq-\lambda/6$ for $\sin\delta\approx1$. We see from Eqs.~(\ref{Jcp1}) and (\ref{Jcp2}) that \CP violation in both lepton and quark sectors characterized by the Jarlskog invariant is correlated, provided that $\lambda,\delta,h,f$ are common parameters to both sectors,
 \begin{eqnarray}
  J^{q}_{CP}=-6hf\lambda^{5}J^{\ell}_{CP}~.
 \end{eqnarray}
This leads to $|J^{\ell}_{CP}|\simeq\lambda^{-5}J^{q}_{CP}\gg J^{q}_{CP}$ from Eq.~(\ref{eq:QMfh}). Equivalently, by using the conventional parametrization of the PMNS matrix~\cite{PDG} and Eq.~(\ref{PMNS}), one can deduce an expression for the Dirac CP phase $\delta_{CP}$:
\begin{eqnarray}
  \delta_{CP}
 &=& -\arg \left(\frac{\frac{U^{\ast}_{e1}U_{e3}U_{\tau1}U^{\ast}_{\tau3}}{c_{12}c^{2}_{13}c_{23}s_{13}}+c_{12}c_{23}s_{13}}{s_{12}s_{23}}\right)~.
\label{DCP}
\end{eqnarray}

\begin{figure}[t]
\begin{tabular}{cc}
\includegraphics[width=7.0cm]{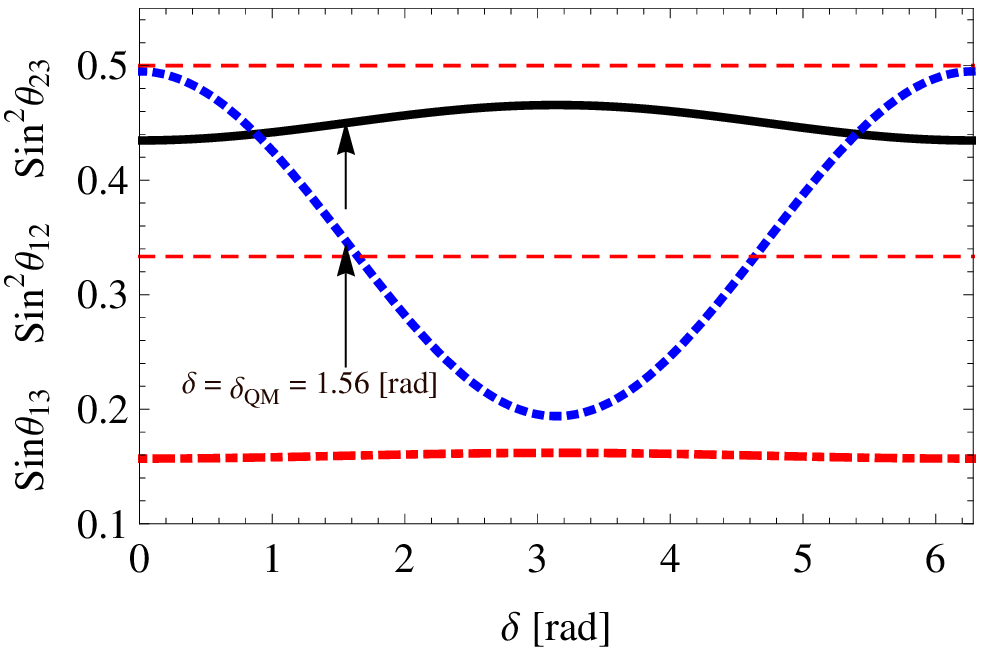}&
\includegraphics[width=6.9cm]{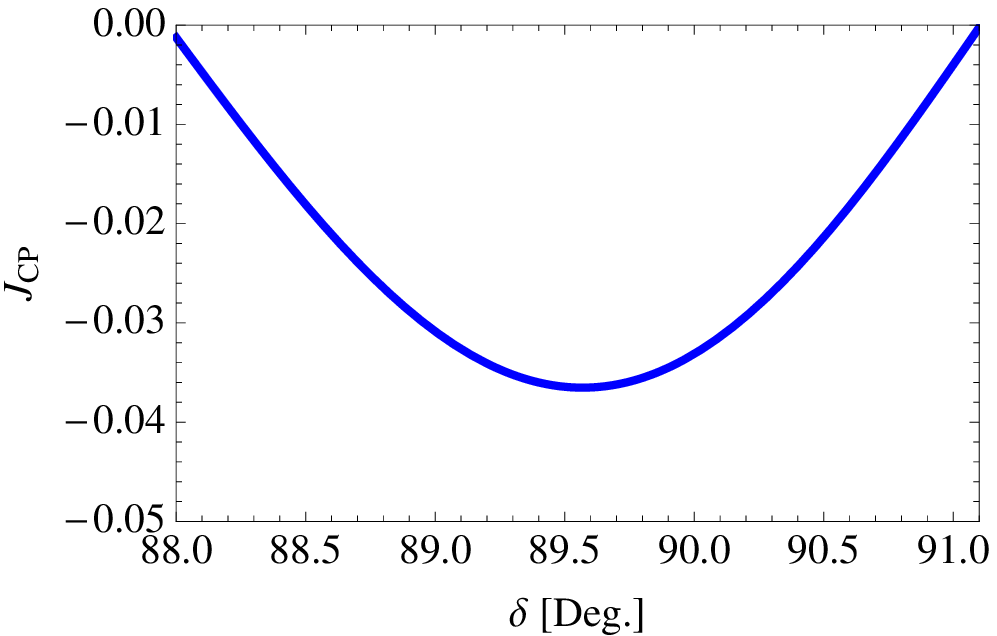}
\end{tabular}
\caption{\label{Fig1} The left plot represents the atmospheric ($\sin^2\theta_{23}$, solid), solar ($\sin^2\theta_{12}$, dotted) and reactor ($\sin\theta_{13}$, dot-dashed) mixing angles as a function of the phase $\delta$, where the horizontal dashed lines denote the TBM  values $\frac{1}{2}$ and $\frac{1}{3}$ for $\sin^2\theta_{23}$ and $\sin^2\theta_{12}$, respectively. The right plot shows the Jarlskog invariant $J_{CP}$ versus the parameter $\delta$. }
\end{figure}

Before proceeding to the numerical analysis,  we exhibit again the experimental data of Eq.~(\ref{exp}) in terms of $\sin^{2}\theta_{12},~\sin^{2}\theta_{23}$ and $\sin\theta_{13}$ at $1\sigma~(3\sigma)$ level:
 \begin{eqnarray}
  \sin^{2}\theta_{12} &=& 0.307^{+0.018~(+0.052)}_{-0.016~(-0.048)}\ , \nonumber\\
  \sin\theta_{13}&=&\left\{
  \begin{array}{ll}
  0.155^{+0.008~(+0.022)}_{-0.008~(-0.025)} & \hbox{NO} \\
  0.156^{+0.007~(+0.021)}_{-0.008~(-0.025)} & \hbox{IO}
  \end{array}
  \right.
  \ ,\qquad\sin^{2}\theta_{23}=
  \left\{
    \begin{array}{ll}
      0.386^{+0.024~(+0.251)}_{-0.021~(-0.055)} & \hbox{NO} \\
      0.392^{+0.039~(+0.271)}_{-0.022~(-0.057)}  & \hbox{IO}
    \end{array}
  \right.
   \ .
 \label{exp1}
 \end{eqnarray}
For the purpose of illustration, we employ the values of the parameters $\lambda,~f,~h$ and $\delta$ given in the quark sector (see Eq.~(\ref{eq:QMfh})). Then we have the predictions
 \begin{eqnarray}
  \sin^{2}\theta_{12}=0.346~,\qquad\sin^{2}\theta_{23}=0.450~,\qquad
  \sin\theta_{13}=0.159~,\qquad J^{\ell}_{CP}\simeq-\frac{\lambda}{6}~,
 \label{prediction}
 \end{eqnarray}
and the corresponding mixing angles are $\theta_{12}=36.0^{\circ},~\theta_{23}=42.1^{\circ}$ and $\theta_{13}=9.2^{\circ}$, respectively. Hence, our prediction for $\theta_{13}$ is consistent with the recent measurement of the reactor neutrino mixing angle.
Fig.~\ref{Fig1} shows the behaviors of mixing parameters as a function of $\delta$ for $\lambda,~f,~h$ having the central values given by Eq.~(\ref{eq:QMfh}).
The left plot of Fig.~\ref{Fig1} represents the atmospheric ($\sin^{2}\theta_{23}$), solar ($\sin^{2}\theta_{12}$) and reactor ($\sin\theta_{13}$) mixing angles as a function of the phase $\delta$, where the horizontal dashed lines denote the TBM values $\frac{1}{2}$ and $\frac{1}{3}$ for $\sin^{2}\theta_{23}$ and $\sin^{2}\theta_{12}$, respectively. As can be seen in this plot,  the deviation of the mixing angles $\theta_{12}$ and $\theta_{23}$ from the TBM pattern in the case of $\delta=\delta_{\rm QM}$ is fairly small: $\frac{1}{2}-\sin^{2}\theta_{23}\simeq0.050$ and $\sin^{2}\theta_{12}-\frac{1}{3}\simeq0.012$, while the reactor angle $\theta_{13}$ is sizable. The right plot of Fig.~\ref{Fig1} shows the Jarlskog invariant versus the parameter $\delta$, where the value of $\delta_{\rm QM}=89.6^{\circ}$ corresponds to $J_{CP}=-0.035$ or equivalently $\delta_{CP}=-87.0^{\circ}$ from Eq.~(\ref{DCP}). Since the dependence of $J_{CP}$ or $\delta_{CP}$ on $\delta$ is very sensitive as one can see from Eq.~(\ref{Jcp2}) or (\ref{DCP}), this is why in the right plot of Fig.~\ref{Fig1} we focus on the range of $\delta$ in the vicinity of $\delta_{\rm QM}$.

\section{Conclusion}
The recent neutrino oscillation data from Daya Bay Collaboration~\cite{An:2012eh} disfavor the TBM pattern at $5.2\sigma$ level, implying a non-vanishing $\theta_{13}$ and giving a relatively large $\sin^{2}2\theta_{13}=0.097$ (best-fit value) corresponding to $\theta_{13}=8.8^{\circ}$. On the theoretical ground, we have proposed a  simple {\it ansatz} for the charged lepton mixing matrix, namely, it has the QM-like parametrization in which the {\it CP}-odd phase is approximately maximal. Then we have proceeded to
study the deviation of the PMNS matrix from the TBM one arising from the small corrections due to the particular charged lepton mixing matrix we have proposed. We have obtained the analytic results for the mixing angles expanded in powers of $\lambda$: the solar mixing angle
$\sin^{2}\theta_{12}\simeq\frac{1}{3}\left(1+2\lambda\cos\delta+\frac{\lambda^2}{2}\right)$, the atmospheric mixing angle $\sin^{2}\theta_{23}\simeq\frac{1}{2}\left(1+{\cal O}(\lambda^{2})\right)$, the reactor mixing angle $\sin\theta_{13}=\frac{\lambda}{\sqrt{2}}[1+{\cal O}(\lambda^{2}\cos\delta)]$ and the Dirac $CP$-odd phase $\delta_{CP} \simeq \delta$. Therefore, while the deviation of solar and atmospheric mixing angles from the TBM values are fairly small, we have found a nonvanishing reactor mixing angle $\theta_{13}\simeq 9.2^{\circ}$ and a very large Dirac phase $\delta_{CP} \simeq \delta_{\rm QM} \simeq {\cal O} (90^{\circ})$. Furthermore, we have shown that the leptonic \CP violation characterized by the Jarlskog
invariant is $|J^{\ell}_{CP}|\simeq\lambda/6$, which could be tested in the future experiments
such as the upcoming long baseline neutrino oscillation ones.

\vskip 1 cm {\bf Acknowledgments}

We are grateful to Zhi-Zhong Xing for useful discussion.
This work was supported in part by the National Science Council of R.O.C. under Grants Numbers:  NSC-97-2112-M-008-002-MY3, NSC-97-2112-M-001-004-MY3 and NSC-99-2811-M-001-038.



\end{document}